\documentclass[fleqn,usenatbib]{mnras}

\usepackage{newtxtext,newtxmath}

\usepackage{caption}
\usepackage{graphicx}

\DeclareCaptionFormat{cont}{#1 (cont.)#2#3\par}
\newcommand{\rthis}[1]{\textcolor{black}{#1}}

\usepackage[T1]{fontenc}

\DeclareRobustCommand{\VAN}[3]{#2}
\let\VANthebibliography\thebibliography
\def\thebibliography{\DeclareRobustCommand{\VAN}[3]{##3}\VANthebibliography}

\usepackage{graphicx}	
\usepackage{amsmath}	
\usepackage{caption}
\usepackage{subcaption}
\usepackage{color}
\usepackage{aas_macros}

\usepackage{lipsum}

\newcommand{\Mpch}{\ensuremath{h^{-1}{\rm Mpc}}}
\newcommand{\Mhsq}{\ensuremath{h^{-2}M_{\odot}}}

\title[Test of constant halo surface density for $\Lambda$CDM]{A test of invariance of dark matter halo surface density using multiwavelength  mock galaxy catalogues}

\author[Gopika K., et al.]{
Gopika K.,$^{1}$\thanks{E-mail: ph19resch01001@iith.ac.in)}
Shantanu Desai,$^{1}$
Aseem Paranjape$^{2}$
\\
$^{1}$Department of Physics, Indian Institute of Technology, Hyderabad, Kandi, Telangana-502284, India\\
$^{2}$Inter-University Centre for Astronomy and Astrophysics, Ganeshkhind, Post Bag 4, Pune 411007,  India\\
}

\date{Accepted XXX. Received YYY; in original form ZZZ}

\pubyear{2023}

\begin{document}
\label{firstpage}
\pagerange{\pageref{firstpage}--\pageref{lastpage}}
\maketitle

\begin{abstract}
A large number of observations have shown that the dark matter halo surface density, given by the product of halo core radius  and core density is nearly constant for a diverse suite of galaxies. Although this invariance of the  halo surface density is violated  at galaxy cluster and group scales, it is still an open question on whether the aforementioned constancy  on galactic scales can be explained within $\Lambda$CDM.   For this purpose, we probe the variation of  halo surface density as a function of mass 
using multi-wavelength mock galaxy catalogs from $\Lambda$CDM simulations, where the adiabatic contraction of dark matter halos in the presence of baryons has been taken into account. We find that  these  baryonified $\Lambda$CDM halos were best fitted with a generalized-NFW profile, and the halo surface density from these halos has a degeneracy with respect to  both the halo mass and the virial concentration. We find that the correlation with mass when averaged over concentration is consistent with a constant halo surface density. However,  a power-law dependence as a function of halo mass also cannot be ruled out.
\end{abstract}




\section{Introduction}

Although, the current concordance $\Lambda$CDM model of cosmology consisting of  70\% dark energy and 25\% cold dark matter is very successful in explaining cosmological observations at large scales~\citep{Planck18}, there are still some  vexing issues with this standard cosmological model. Some of these problems with the standard model include the core-cusp and the  missing satellites problem~\citep{Bullock}, Hubble constant tension~\citep{Divalentino}, $\sigma_8$  tensions between cluster and CMB~\citep{Benisty}, failure to detect cold dark matter candidates in laboratory based experiments~\citep{Merritt}, Lithium-7 problem in Big-Bang nucleosynthesis~\citep{Fields}, radial acceleration relation in spiral galaxies with very low scatter~\citep{McGaugh16},   CMB anomalies at large angular scales~\citep{Copi},  etc. An up-to-date  summary of the  challenges for $\Lambda$CDM model can be found in ~\citet{Periv,Abdalla22,Peebles22}.  Therefore a large number of alternatives  to the standard cosmological model  have been proposed to account for some of these anomalies~\citep{alternatives,Banik,Abdalla22}.

An intriguing observational result discovered more than a decade ago is that the dark matter halo surface density  was found to be constant,  for a wide variety of systems spanning over 18 orders in blue visual magnitude  for a diverse suite of galaxies, such as  spiral galaxies, low surface brightness galaxies,  dwarf spheroidal satellites of Milky way~\citep{Donato04,Donato,Walker,Salucci,Kormendy14,Chiba,Burkert15,Salucci19}. These results for a constant dark matter surface density were obtained by fitting the dark matter distribution in these systems to  a cored profile, either  Burkert~\citep{Burkert95}, pseudo-isothermal profile~\citep{Kormendy14}, or a simple isothermal sphere~\citep{Spano}. All these cored profiles can be parameterized by a  central density ($\rho_c$) and core radius ($r_c$); and the halo surface density ($S$) is then defined according to:
\begin{equation}
 S \equiv \rho_c \times r_c    
\label{eq:halosurfacedensity}
\end{equation}

The halo surface density can also be interpreted as  a measure of the Newtonian acceleration at a given radius~\citep{Cardone}.
The existence of a constant dark matter surface density was found to be  independent of which cored profile was used~\citep{Donato}. The latest best-fit value obtained by fitting all the available data for the above systems is given by $\log(\rho_c r_c) = 2.15 \pm 0.2$ with units of $\log (M_{\odot}/pc^2)$~\citep{Salucci19}.

However, cored profiles cannot adequately fit all systems, some of which need cuspy profiles~\citep{Simon,Strigari,Genina,Helmi,Newman,Newman2,Hayashi20}. Also, sometimes the same data can be easily fitted with both cored and cuspy profiles~\citep{Salucci07,Meadows,Boldrini}. 
Therefore, a variant of the aforementioned halo surface density  has been constructed (which has been refered to in the literature as ``column density''),  whose value was found to be  nearly invariant for both cored and cuspy profiles~\citep{Boyarsky,BoyarskyPRL,DelPopolo12}, and therefore can be applied to observational data,  since it is mostly agnostic to the underlying dark matter profile:
\begin{equation}
S (R)  = \frac{2}{R^2} \int_0^{R} r' dr' \int_{-\infty}^{+\infty} dz \rho_{DM} (\sqrt{r'^2+z^2})
\label{eq:SB}
\end{equation}
The above definition implies that $S(R)$  is proportional to the dark matter surface density at a radius $R$.
The column densities for  some of the commonly used dark matter profiles such as Navarro-Frenk-White (NFW) ($S_{NFW}$), pseudo-isothermal ($S_{ISO}$) and Burkert profiles ($S_{BUR}$) are connected to each other as follows: $S_{NFW} (r_s) \approx  0.91 S_{ISO} (6 r_c) $ and $S_{NFW} (r_s)  \approx 0.98  S_{BUR} (1.66 r_0)$~\citep{Boyarsky}. The analytical expressions for  $S_{NFW}$ and $S_{ISO}$  have been provided in ~\cite{Boyarsky,DelPopolo12}~.\footnote{We note that there are also  other variants of the halo surface density used in the literature, for example mean surface density, given by $\frac{M(r)}{4/3 \pi r^2}$~\citep{Bondarenko} or $\frac{M(r)}{\pi r^2}$~\citep{Chiba}. However, we do not evaluate these quantities in this work.}

The aforementioned results for the constancy of dark matter halo surface density have however been  disputed \rthis{in the literature by many other works},   which have shown that the dark matter column density (Eq.~\ref{eq:SB}) as well as the halo surface density (Eq.~\ref{eq:halosurfacedensity}) is not constant and is correlated with the galaxy luminosity,  halo mass, and stellar age~\citep{Boyarsky,Napolitano,Cardone, DelPopolo12,Cardonedel,Saburova,DelPopolo17,DelPopolo20}. Some of these works have found $S \propto M_{halo}^{0.16-0.2}$~\citep{DelPopolo12} as well as  $S \propto L^{0.13}$~\citep{DelPopolo20}. \rthis{ Most recently, ~\citet{DelPopolo23} found a dichotomy in the behaviour of halo surface density based on the effective surface brightness ($\Sigma_{\rm eff}$). For high-surface brightness galaxies from the SPARC sample,  ($\Sigma_{\rm eff}>200 L_{\odot}/pc^2$), this work found the halo surface density is constant as a function of magnitude, but with  a larger value than ~\citet{Donato}. However for low-surface brightness galaxies, the halo surface density is anti-correlated with magnitude  or correlated with luminosity~\citep{DelPopolo23}.
Therefore, a majority of papers in the literature could not corroborate the claims in ~\citet{Donato}.}

These observations of a (near)-constancy of the halo surface density have been shown to be a very good laboratory for testing alternatives to $\Lambda$CDM or constraining particle dark matter properties within these alternatives.~\citet{Milgrom} has pointed out that, in MOND, the numerical value of the  constant halo surface density obtained in ~\citet{Donato} is equal to $\frac{a_0}{2\pi G}$, where $a_0$ is the MOND acceleration scale, and hence this observation is consistent with MOND predictions. A constant halo surface density is also predicted by superfluid dark matter~\citep{Berezhiani}, self-interacting dark matter~\citep{Rocha,Bondarenko,Loeb,Kaplinghat15} and ultra-light scalar dark matter~\citep{Arturo}. Observations of constant surface density have been used to obtain a lower limit on the warm dark matter particle mass~\citep{Chiba}.  \citet{Burkert20} has argued that a constant halo surface density is also in tension with fuzzy dark matter. However, the analysis in ~\citet{Burkert20}  assumes that the core density of fuzzy dark matter models follows isothermal or Burkert profile, whereas the density profiles in these models are quite different~\citep{Khelashvili}.   Alternatives to $\Lambda$CDM model such as the logtropic model, which involve coupled dark energy-dark matter have also been able to explain this constancy of halo surface density~\citep{Chavanis}.

Motivated by these considerations, studies on the constancy of  halo surface density were extended to galaxy cluster~\citep{Chan,Gopika_2020} as well as galaxy group scales~\citep{Gopika2021}. These works found that although the core radius is inversely proportional to the  core density, their product, viz. the halo surface density is about an order of magnitude larger than that obtained for single galaxies, whereas for galaxy groups, it is in between that for single galaxies and groups.  Therefore, the invariance of halo surface density  found from galaxies~\citep{Donato} does not hold true for group or cluster scaled halos.

It is still an open question whether the current standard model of Cosmology, namely the $\Lambda$CDM model, with best-fit parameters determined by Planck Cosmology~\citep{Planck18}, can explain the (near) constancy of the  halo surface density on galactic scales and its elevated value on cluster scales. Some proposed semi-analytical mechanisms within the $\Lambda$CDM model to explain the near constancy of halo surface density include transformation of cusps to cores due to dynamical feedback processes~\citep{Mori}, non-violent relaxation of galactic halos~\citep{Baushev}, secondary self-similar infall model~\citep{Delpopolo09,DelPopolo12}, analytical description of the virialized spherical dark matter halo with the NFW density profile~\citep{Demianski}. However, except for the secondary infall model, which predicts ($S \propto M_{200}^{0.20 \pm 0.05}$) and the ~\citet{Demianski} model which predicts $S \propto M_{vir}^{0.3}$, none of the other proposed models can self-consistently predict the observed trends for halo surface densities  as a function of mass for the full dynamical range of halo mass from dwarf galaxies to cluster scale halos. 
More than a decade ago, a preliminary study using isolated halos and subhalos from the Aquarius cosmological simulations~\citep{Dutton,Springel08} also found a scaling of the column density with halo mass, although no error analysis was carried out~\citep{Boyarsky}.
In the last few years the availability of high resolution mock catalogs, which agree with observational data enables a detailed systematic study of the dark matter surface density.

Therefore, in this work we set out to carry out such a study using mock galaxy catalogs~\citep{Paranjapemock} that successfully mimick the properties and spatial distribution of galaxies in the local Universe (redshift $z\lesssim0.1$). Among other features, these mock catalogs have been shown by \citet{ps21} to reproduce the radial acceleration relation~\citep{McGaugh16}, which has previously been an enigma within  the $\Lambda$CDM model~\citep{Famaey12}.

This manuscript is structured as follows. The multi-wavelength mock galaxy catalog is defined in Sect.~\ref{sec:mock}. The data analysis procedure is described in Sect.~\ref{sec:analysis}. The gNFW fits are described in Sect.~\ref{sec:gNFWfits}. Our results for the halo surface density can be found in Sect.~\ref{sec:results}. We conclude in Sect.~\ref{sec:conclusions}.

\section{Mock galaxy catalogs}
\label{sec:mock}
In this work, we use mock galaxy catalogs constructed using the algorithm presented by  \citet{Paranjapemock}. Here, we briefly summarise the main features of interest of these catalogs and refer the reader to \citet{Paranjapemock} for further details of the underlying $N$-body simulation and the mock algorithm which was extensively tested by them. 

Each catalog represents a luminosity-complete sample of galaxies with SDSS $r$-band absolute magnitude $M_r\leq-19$ in a $(300\Mpch)^3$ comoving volume at $z=0$. The mock contains both central and satellite galaxies, populated in dark matter  halos identified in an $N$-body simulation having $1024^3$ particles with a flat $\Lambda$CDM WMAP7 cosmology (\citealp{komatsu+2011}; $\Omega_{\rm m}=0.276$, $h=0.7$). 

The galaxies in each mock were populated using a halo occupation distribution (HOD) model  calibrated by \citet{ppp19} using luminosity- and colour-dependent clustering measurements from the Sloan Digital Sky Survey \citep[SDSS,][]{york+00,zehavi+11}. Each galaxy in the mock is assigned absolute magnitudes in SDSS $u$, $g$ and $r$ bands and a stellar mass $m_\ast$ using a colour-dependent mass-to-light ratio. 

In this work, we use the population of mock central galaxies having $m_\ast\geq10^{9.8}\Mhsq$, which results in $\sim221,000$ objects that form a volume-complete sample of stellar mass-selected galaxies.

The host halos of these centrals are `baryonified' as described by \citet{Paranjapemock}, and the dark matter profiles of the respective host halos are assumed to respond to the presence of these baryons. This response is modelled as a quasi-adiabatic relaxation process \citep[][see section~3 of \citealp{ps21}]{teyssier+11,st15} parametrised by a `relaxation parameter' $q$ to  control the amount of relaxation, such that $q=1$ ($q=0$) corresponds to perfect angular momentum conservation (no response to baryons). We have performed our analysis using two mock catalogs, respectively produced using the values $q=0.68$ 
and $q=0.33$. While the former provides a good description of the effect seen in cluster-sized halos in hydrodynamical CDM zoom simulations \citep{teyssier+11,st15}, it is subject to some theoretical uncertainty as discussed by \citet{ps21}. The second value $q=0.33$ was shown by \citet{ps21} to improve the agreement at high accelerations between the median RAR of the mock and measurements using disk galaxies in the SPARC sample. Settling the theoretical uncertainty in the modelling of quasi-adiabatic relaxation requires a detailed study of hydrodynamical simulations of galaxy formation over a large mass range, which has only recently begun to be performed \citep{vp23}. We have found, however, that all our results in this work are independent of the value of $q$. For brevity, therefore, we only report the results using $q=0.33$.

\section{Data Analysis}
\label{sec:analysis}
 In this section, we will outline our data analysis procedure. Each simulated halo is characterized by a  baryonified dark matter density profile up to the virial radius, along with other physical parameters such as  the virial concentration ($C_{vir}$), virial-mass ($M_{halo}$), and stellar mass ($m_\ast$). Firstly, we shall describe the  different parametric forms of the dark matter halo density profiles used to fit the simulation data. Then, we discuss   the regression method  used to determine the best-fit density profile, which can robustly  fit the full range of  simulated  profiles. Lastly, we shall present our results for the halo surface density using the best-fit profile and study its variation as a function of halo mass and other parameters.
 \subsection{Dark Matter Density Profiles}
 \label{sec:analysis1}

All  the results on the halo surface density in literature were obtained by  fitting for    cored profiles such as  isothermal or  Burkert profiles~\citep{Donato, Salucci, Chan, Gopika_2020, Gopika2021, DelPopolo20} or NFW profiles~\citep{Boyarsky, BoyarskyPRL}. As a first step towards this,  we produce density profiles of relaxed dark matter $\rho(r)$ in  units of $\rho_{vir}$ at 30 logarithmically spaced values of $r/R_{vir}$ in the range $(10^{-5}, 1)$. These densities and radii are then converted into the real physical units of  $M_{\odot}/pc^3$ and $kpc$ respectively by assuming the fiducial cosmology of the underlying simulation.

\begin{figure}
    \centering
	\includegraphics[width=\columnwidth]{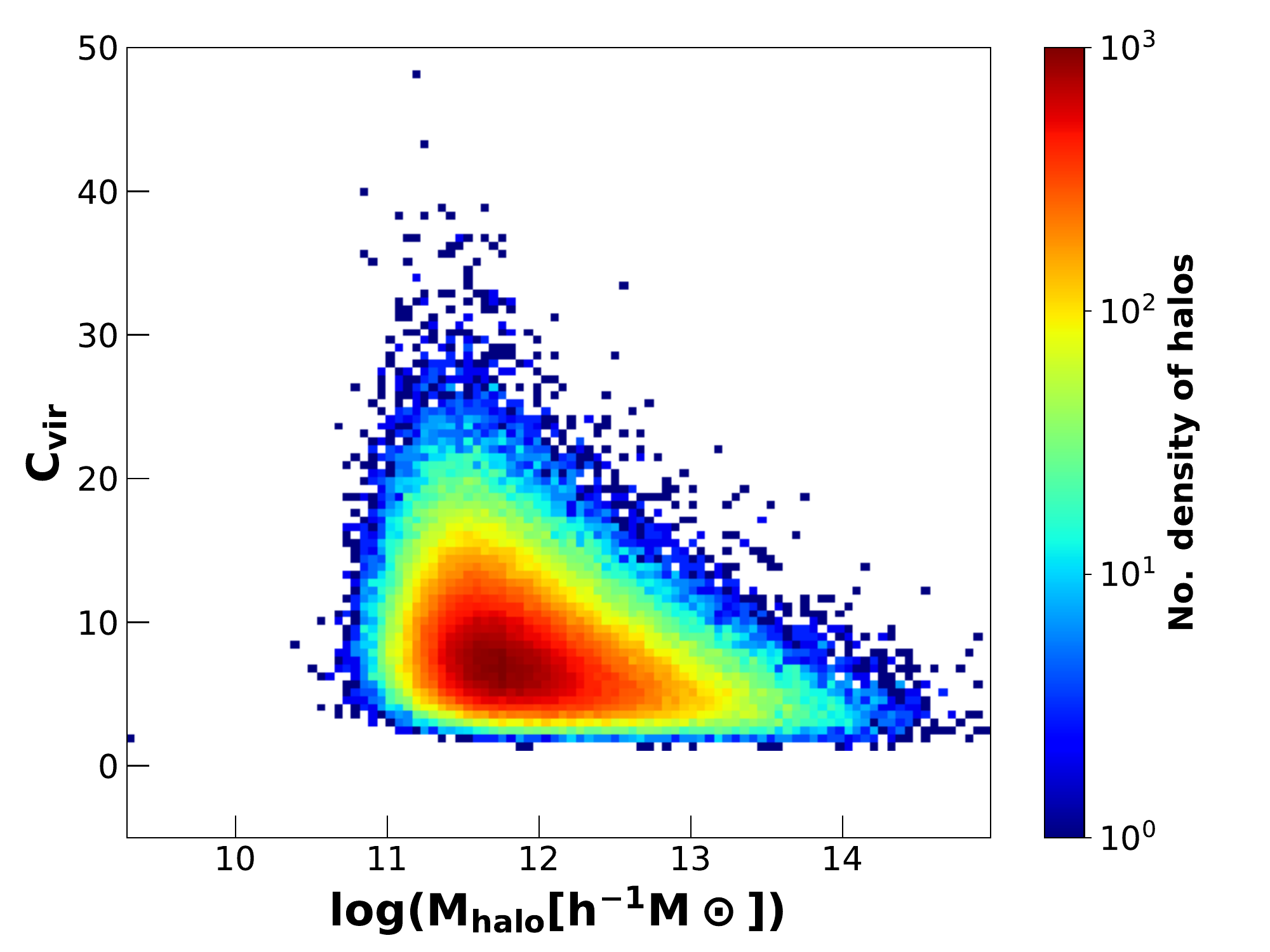}
    \caption{Distribution of the virial mass, $M_{halo}$ and virial concentration, $C_{vir}$ for the stellar-mass complete sample of simulated halos.}
    \label{fig:figure1.1}
\end{figure}

 Figure~\ref{fig:figure1.1} shows the distribution of the halo mass and concentration for the stellar-mass complete sample of 2,21,670 halos with a stellar mass selection, $\log m_\ast\geq9.85h^{-2}M_\odot$ included in this analysis. We have used the total mass at the virial radius $r_{200}$ as a proxy for the halo mass($M_{halo}$) of the dark matter halos for the entirety of this study.

We now fit the simulated halos with a cored isothermal and Burkert profile as well as  some non-cored profiles such as  NFW and generalized-NFW (gNFW), which are described below:

\begin{itemize}
\item Burkert Profile~\citep{Burkert95}
 \begin{equation}
\rho_{Bur}(r) =\frac{\rho_c r_c^{3}}{(r^2+r_c^2)(r+r_c)}
\label{eq:eqbur} 
\end{equation}
\item Isothermal Profile~\citep{Kormendy14}:
 \begin{equation}
\rho_{iso}(r) =\frac{\rho_c}{[1+(r/r_c)^2]}
\label{eq:eqiso} 
\end{equation}
where $r$ is the radial distance; $r_c$ and $\rho_c$ are the scale parameters, which are usually referred to in the literature as  the core radius and core density, respectively.
\item NFW Profile~\citep{NFW}:
 \begin{equation}
\rho_{NFW}(r) =\frac{\rho_s}{(r/r_s)(1+r/r_s)^2}
\label{eq:eqnfw} 
\end{equation}
where $r_s$ and $\rho_s$ are the NFW scale parameters.
\item gNFW Profile~\citep{Zhao1996,Moore}:
 \begin{equation}
\rho_{gNFW}(r) =\frac{\rho_s}{(r/r_s)^\alpha(1+r/r_s)^{3-\alpha}}
\label{eq:eqgnfw} 
\end{equation}
This 3-parameter gNFW model is a generalization of the NFW profile where $\rho_s$ and $r_s$ are the normalization and scale radius, respectively and $\alpha$ is the inner slope of the profile, where $\alpha = 1$ would attain a standard NFW profile.  
\end{itemize}

However, we shall subsequently show  that only a generalized NFW profile can holistically fit  these baryonified dark-matter halos for all galaxies as discussed in the next section. 

\subsection{Fits to halo profiles}
\label{sec:analysis2}
The fits for the  simulated density profiles to the analytic density models described in Section~\ref{sec:analysis1} are paramount to  this study.  Here, we have  performed fits on the stacked halo profiles. Stacked profiles are essential, since the individual halo profiles do not have any errors associated with them, and hence analysing the stacks permits us to average over a bin of data to obtain  a single density profile with error proxies  derived from looking at the scatter within the stack. The use of stacks in this work is adequate, as  our main goal is to study the halo scaling relations. 

For the $\chi^2$ analysis associated with the fitting of the simulated halo densities with cored isothermal, NFW, and generalized-NFW models, we have considered the distributed densities for fixed bins of halo-mass ($M_{halo}$), virial concentration ($C_{vir}$), and stellar mass ($m_\ast$). For this work, we have used three different binning procedures to achieve three distinct  data sets labelled as Set 1, Set 2, Set 3, which correspond to  binning the entire simulation data with respect to $M_{halo}$; $M_{halo}$ and $C_{vir}$; and  $M_{halo}$, $C_{vir}$, and $m_\ast$, respectively with  a very stringent technique being followed in order to attain evenly populated bins for each of  these cases. 

For the Set 1 data, the whole population is divided into five equal quantiles of halo mass with each mass-bin having approximately 44,300 halos. Additionally, in order to carry out  a more detailed analysis,  we can perform a similar percentile binning with respect to the  virial concentration in the Set 1 data to get Set 2 data. Each bin of mass is further sub-divided using  $C_{vir}$ giving 25 bins, having an approximate population of 8800 halos in each bin. The Set 3 data is similarly obtained by doing a percentile binning in the Set 2 data using the stellar mass to get an evenly populated 125 three-dimensional bins in  $M_{halo}$, $C_{vir}$, and $m_\ast$. 

A stack can be constructed by calculating the mean density of the simulated profiles in each of the percentile-bins discussed above for the 30 logarithmically spaced values of $r$ in the range $(10^{-5}, 1)\times R_{vir}$. The error associated with the density in each bin is represented by the standard deviation of the density of the given radial bin distributed across all halos of the stack. The physical parameters such as $M_{halo}$, $C_{vir}$, and $m_\ast$ associated with these stacked objects (for each bin)  are obtained from the median of the binned data. 

\begin{figure}
    \centering
	\includegraphics[width=\columnwidth]{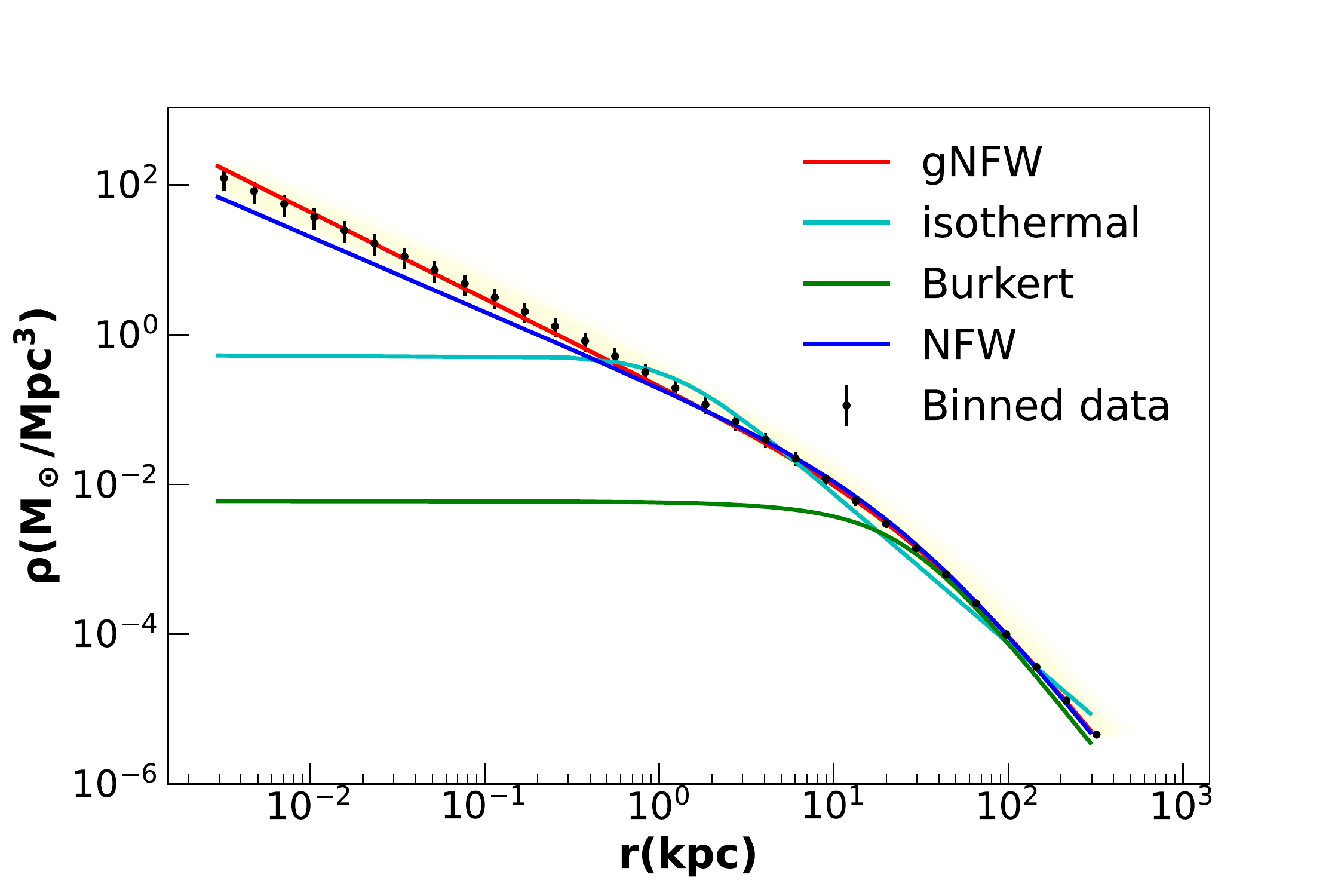}
    \caption{This figure shows the  relaxed dark-matter density profile for a stacked object with $M_{halo}=2.03\times 10^{12}h^{-1}M_{\odot}$,  $C_{vir}=9.83$  and $\log m_\ast=2.47\times 10^{10}h^{-2}M_{\odot}$ which is fitted with a generalized-NFW (red), NFW (blue), isothermal (cyan) and Burkert(green) profiles. The yellow region shows the distribution of halo profiles in the mass range within this bin.}
    \label{fig:figure1}
\end{figure}

Finally, the stacks are each fitted to the functional forms of all the halo profiles, to determine the best-fit parameters  in each model followed by a model selection procedure to determine the best-fit parametric model which can fit the full range of  halo density profiles.
In Figure \ref{fig:figure1}, a radial profile of the relaxed dark matter density for one of  the Set 3 stacked objects is shown. The data points represent the average density profile of a bin where the data are binned with respect to the halo mass, $C_{vir}$, and stellar mass. The median values of the physical parameters $M_{halo}$, $C_{vir}$, and $m_\ast$ associated with this stacked halo  are $2.03\times 10^{12}h^{-1}M_{\odot}$, 9.83, and $2.47\times 10^{10}h^{-2}M_{\odot}$, respectively.  Based on this visual appearance, we can see that only the gNFW profile provides a good fit. However, we quantify this using  model comparison techniques as described in the next subsection.

\subsection{$\chi^2$- Analysis and Model Comparison}
\label{sec:analysis3}
\begin{figure}
    \centering
	\includegraphics[width=\columnwidth]{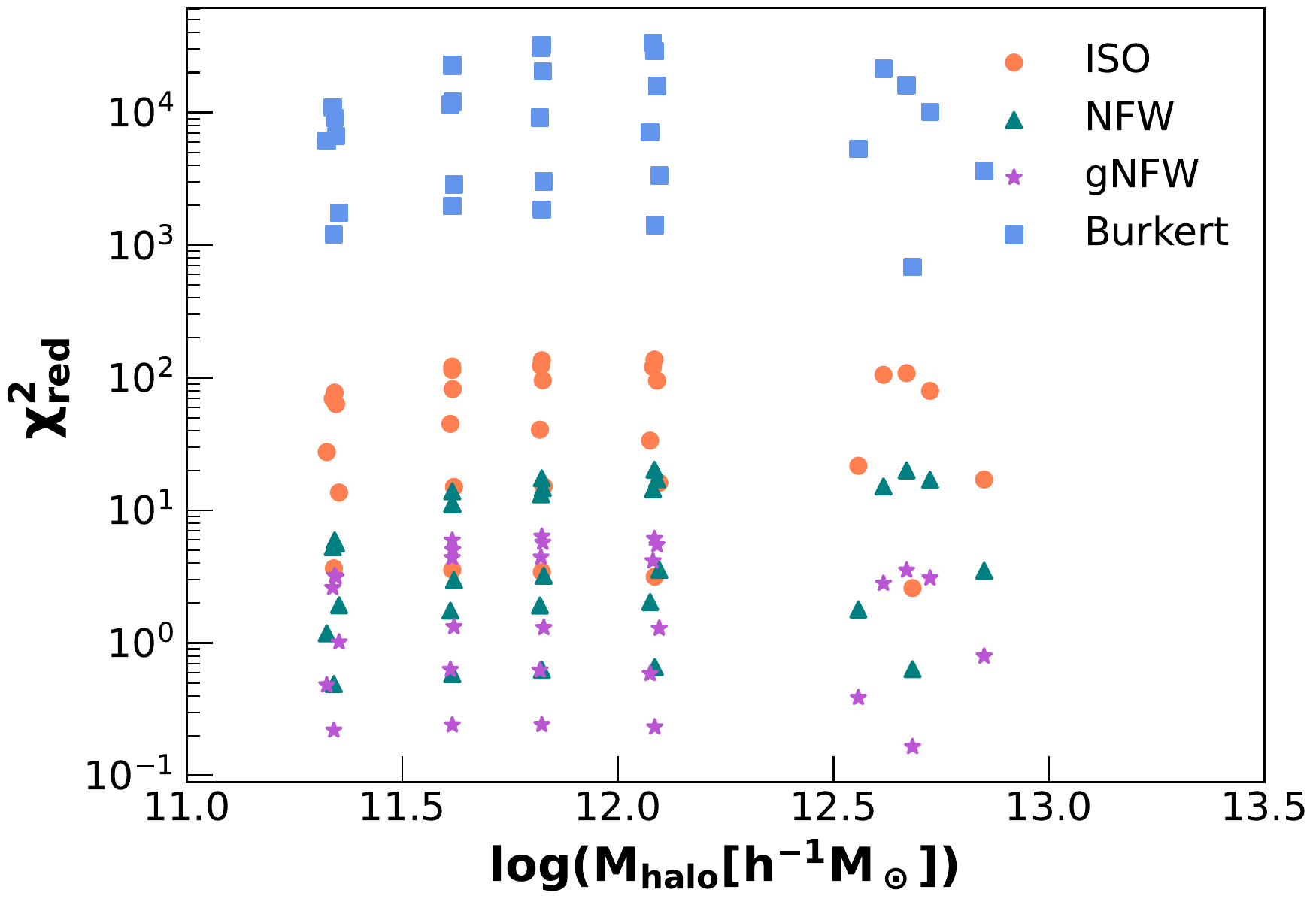}
    \caption{Distribution of the reduced-$\chi^2$ values of the fits for isothermal (orange), NFW (teal), gNFW (violet) and Burkert (blue) profiles to the stacked data across the halo mass. }
    \label{fig:figure1.2}
\end{figure}

The efficacy of the dark matter analytic model fits to the simulation data discussed in Section \ref{sec:analysis2} has been gauged using the $\chi^2$ analysis supplemented with information theory techniques such  as Akaike and Bayesian information criteria~\citep{Liddle07}. The $\chi^2$ statistics is first used to determine the goodness of fit for the aforementioned models. We have  estimated the reduced $\chi^2$ for the fits which is given by,
\begin{eqnarray}
    \rm{\chi^2} = \sum \bigg(\frac{\rho_{model}-\rho_{data}}{\sigma}\bigg)^2  \quad  and \quad \rm{\chi^2_{red}} = \frac{\chi^2}{N-k} 
\end{eqnarray}
Here, $\rho_{model}$ is the density profile derived from the best fit values for the analytical models used to fit the simulated stacks, whereas  $\rho_{data}$ and $\sigma$ denote the data-points and associated errors of the stacks from simulated halo density profiles respectively, as described in Section \ref{sec:analysis2}, $N$ is the total number of data points, $k$ is the number of free parameters in each model, where the summation is done over all the data points. 
Figure \ref{fig:figure1.2} represents the distribution of reduced-$\chi^2$ values for the binned data. We infer that cored profiles such as isothermal or  Burkert provide egregious fits, whereas  the NFW profile  with reduced $\chi^2$ greater than one for many halos also does not  provide a good description for the full range of simulation data. However, we find that only the  generalized NFW profile can self-consistently fit these baryonified dark-matter halos with reduced-$\chi^2$ close to one. Regardless, drawing a conclusion solely from the reduced-$\chi^2$ values is not advisable as the models described here do not have the same  number of free parameters. Hence to assess the robustness of gNFW fit relative to NFW,  we have used the Akaike Information Criterion (AIC) and the Bayesian Information Criterion (BIC)  metrics to quantify the goodness of fit. The AIC and BIC are defined as follows~\citep{Liddle07}:
\begin{eqnarray}
\rm{AIC} &=&  \chi^2 + 2k + \frac{2k(k+1)}{N-k-1} \\
\rm{BIC} &=& \chi^2 + k \ln N, 
\end{eqnarray}
\begin{table}
\centering
\begin{tabular}{ccccc}
\hline
\hline
\multicolumn{1}{c}{\textbf{Mass-bin}} & \multicolumn{2}{c}{\textbf{gNFW}} & \multicolumn{2}{c}{\textbf{NFW}}\\
\textbf{} &  {AIC} & {BIC} &  {AIC} & {BIC} \\
\hline
\hline
Bin 1  & -127.5 & -124.2 &  -122.1 & -119.7 \\
Bin 2  & -135.3 & -131.9 &  -127.8 & -125.5 \\
Bin 3  & -140.6 & -137.3 &  -131.9 & -129.6 \\
Bin 4  & -146.1 & -142.8 &  -136.4 & -134.0 \\
Bin 5 & -156.1 & -152.8 & -145.2 & -142.9  \\
\hline
\end{tabular}
\caption{\label{tab:aicbic} This table shows the AIC-BIC values for both the gNFW and NFW fits to the mass-binned stacks(Set 1) of the simulated halos.  Since the fits for the cored profiles were bad, we did not apply these AIC/BIC techniques to them. Both AIC and BIC point to gNFW being the preferred profiles for all the bins}
\end{table}
When comparing two models, the model with the smaller value of AIC and BIC is considered the favored one, and a difference of more than ten in AIC/BIC is considered as decisive evidence in favor of the model with the lower value. We have here conducted these model comparison tests for the three-parameter gNFW profile which provides very good fits to all the dark matter halos than the 2-parameter NFW profile. We do not apply the AIC and BIC based tests to the  isothermal and Burkert profiles,  as these models can be excluded based on their large reduced $\chi^2$ values. The AIC and BIC values for the mass-binned data are tabulated in Table~\ref{tab:aicbic} and we see the values for gNFW model  are consistently smaller than the NFW model for all the bins with the difference equal to 5-10 depending on the mass bin.

Therefore, since  the gNFW profile is the most robust in terms of describing the simulated halo density profiles based on the aforementioned  statistical tests,  we shall derive all the other results based on the fits from this profile for the remainder of this paper. 

\section{GNFW fits to   Simulated halos}
\label{sec:gNFWfits}
In this section, we shall discuss the best-fit gNFW parameters  to the stacks obtained  from the mock catalogs. The density data  for one such stacked profile as shown in Figure ~\ref{fig:figure1} is fitted with a gNFW model (Eq.~\ref{eq:eqgnfw}) by maximizing the log-likelihood function using the emcee MCMC sampler~\citep{emcee}. The corner plots of the fit parameters associated with the gNFW model for the mass binned data showing 68\%, 90\%, and 99\% credible intervals can be found in the Appendix.
Figure~\ref{fig:figure2} shows the distribution of the fit parameters of a gNFW model for the Set 2 data, where the halos are binned with respect to  both the halo mass and the  concentration. In this figure, the trend of the gNFW parameters or a combination of them with the halo mass is depicted with the five quantiles of the virial concentration represented using  five distinct symbols. The scale density ($\rho_s$) decreases with halo mass but increases with $C_{vir}$, whereas $r_s$ shows an antipodal trend with the halo mass and $C_{vir}$ to that of the density. 
The near-unity values of the inner-slope ($\alpha$) suggest an inner profile similar to that of the standard NFW halo, which is further compounded by having the logarithmic slope nearly equal to -2 for this model as is shown in Figure \ref{fig:figure2}.

\begin{figure*}
    \centering
	\includegraphics[width=\linewidth]{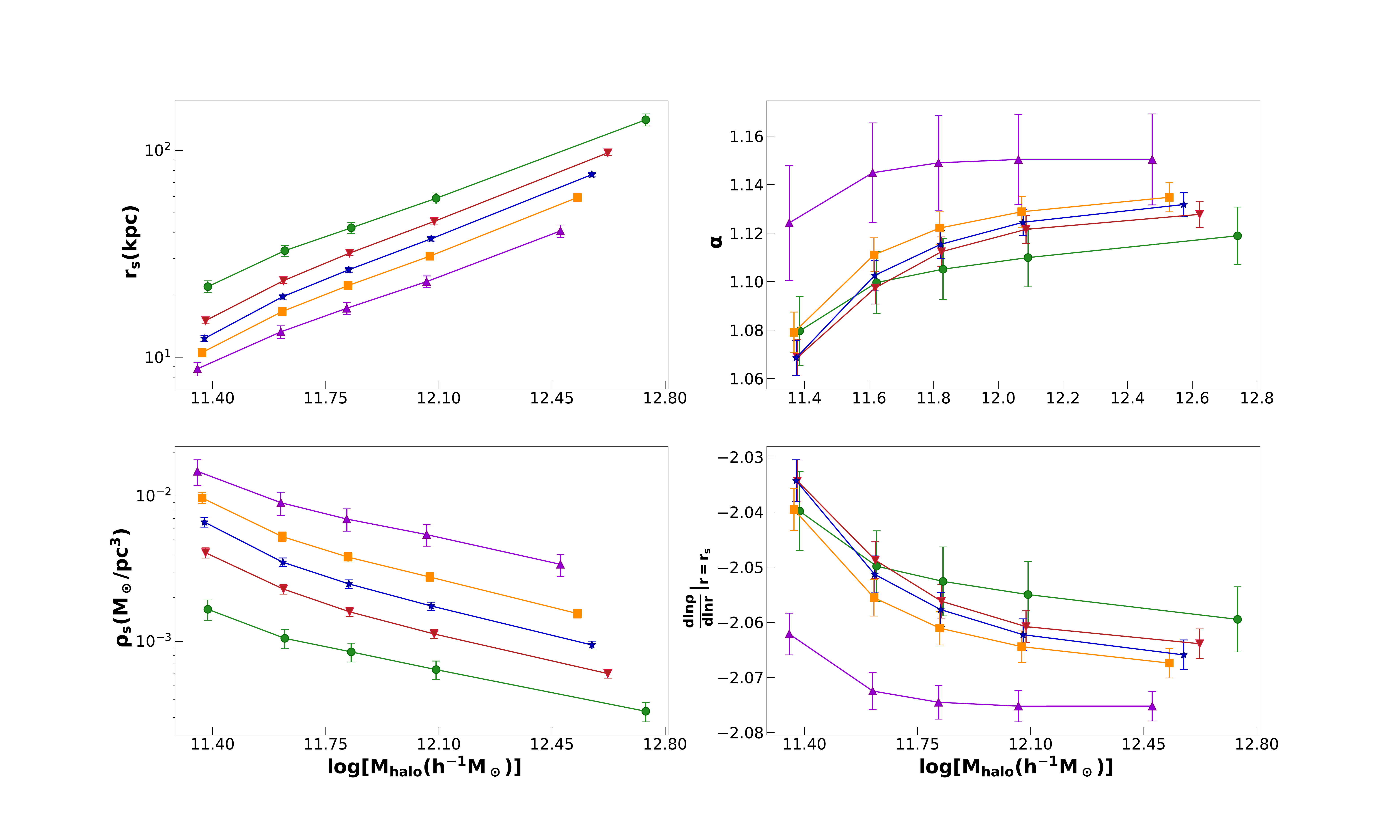}
    \caption{The distribution of best-fit parameters of the gNFW profile as a function of halo mass. The trend of the gNFW parameters with mass  for different $C_{vir}$ bins are designated using different colours. Bins 1, 2, 3, 4, and 5 are represented by green, red, blue, yellow, and violet colours, respectively.}
    \label{fig:figure2}
\end{figure*}

\subsection{Scaling relation of $r_s$  with halo parameters}
\begin{figure*}
    \centering
	\includegraphics[width=0.9\linewidth]{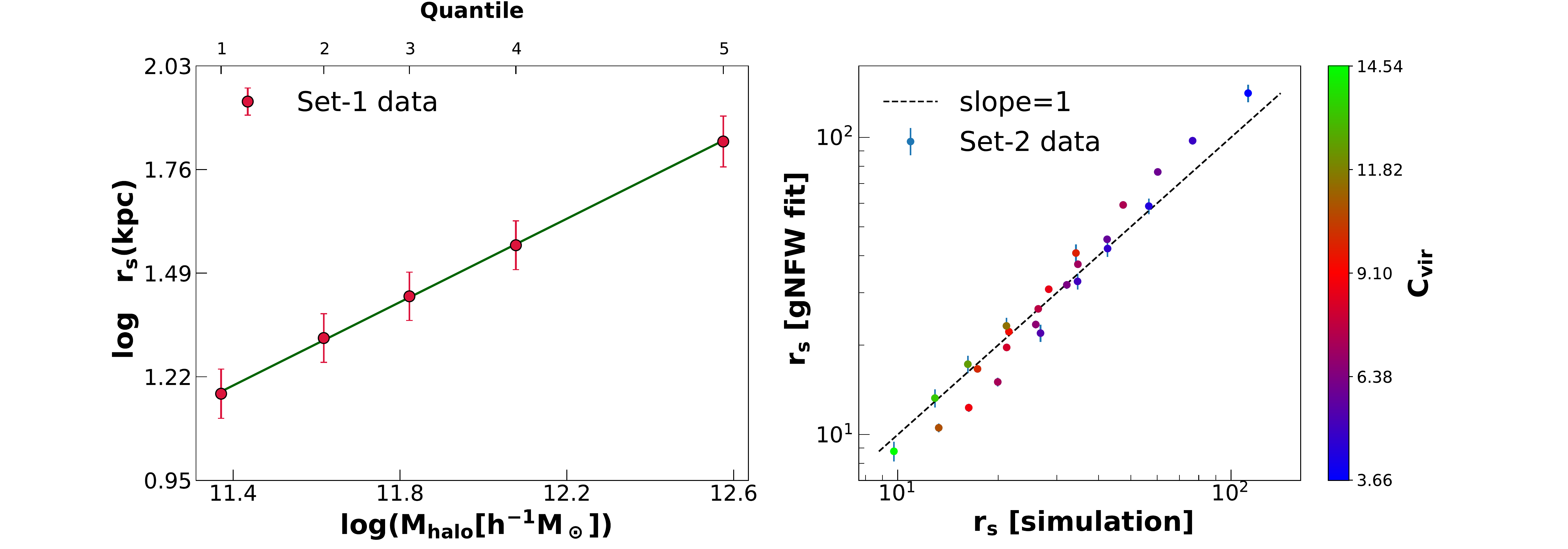}
         \caption{Plot of gNFW scale-radius  ($r_s$) as a function of  halo mass in log-space for the simulated data averaged over equal mass quantiles for Set 1 data (left panel).The gNFW scale radius versus the scale radius derived from the simulation parameters $C_{vir}$ and $M_{halo}$ in log-space for the Set 2 data (right panel). The $r_s$ parameter in both X- and Y-axis is depicted in the physical units of kpc. The colour bar in  the right panel shows the virial concentration associated with the simulated halos. }
    \label{fig:figure3}
\end{figure*}

The mock catalogs discussed in Section \ref{sec:mock} assume a $\Lambda$CDM cosmology, and each simulated halo starts with its total matter density described with an NFW profile that is consistent with the gravity-only simulation, where  this mock catalog is being generated with mass $M_{halo}=(4\pi/3) r_{vir}^3\times 200\rho_{crit}$ and $C_{vir}=r_s(NFW)/r_{vir}$ and later undergoing a baryonification. These modified halos are better described by a generalized NFW profile rather than with the NFW model,  as shown in Section \ref{sec:analysis}. The right panel of Figure~\ref{fig:figure3} shows the distribution of the scale-radius ($r_s(gNFW)$) derived from the gNFW fits to the Set 2 data with their NFW scale radii ($r_s(NFW)$) calculated from the simulation parameters, $C_{vir}$ and $M_{halo}$, which shows a tight positive correlation with an approximate slope of unity, which implies that  there is no considerable change between the gNFW fit parameter, $r_s(gNFW)$ and that derived from the initial simulation parameters. 
This is further compounded by the trend of the logarithmic slope of $\rho_s$ with $M_{halo}$ that has values very close to -2  and depicts a slightly negative trend with $M_{halo}$ and $C_{vir}$ (cf. Figure ~\ref{fig:figure2}). The tight positive scaling relation of $r_s(gNFW)$ with $M_{halo}$ in log-space for the Set 1 data presented in Figure \ref{fig:figure3} (left panel) is given by,
\begin{equation}
    \log r_s = (0.54\pm0.07) \log M_{halo} -(4.96 \pm 0.84)
    \label{eq:rsmass}
\end{equation}
From Eq.~\ref{eq:rsmass}, we infer  that the baryon-affected halos have their $r_s$  increasing with mass slightly faster than $M^{1/3}$, and $C_{vir}$ is marginally decreasing with mass similar to the standard NFW halos.

\section{Results for halo density}
\label{sec:results}
Although the earliest studies of dark matter surface density were done using the halo surface density defined in Eq.~\ref{eq:halosurfacedensity}~\citep{Donato}, the dark matter column density  (Eq.~\ref{eq:SB}) was found to be invariant with respect to the posited  dark matter density profile and  hence, subsequently applied to observational data for NFW~\citep{Boyarsky,BoyarskyPRL,DelPopolo12,DelPopolo17}
 as well as Burkert profile~\citep{Gopika2021}. 
Since neither the  NFW nor any of the  cored profiles could adequately fit the density profiles of the simulated halos, we calculate the dark matter column density   using  the  three-parameter gNFW profile and use that as a proxy for the dark matter halo surface density. The dark matter column density for a gNFW profile is given by~\citep{Boyarsky}:
\begin{equation}
\begin{aligned}
S_{gNFW} (R)  = \frac{2}{R^2} \int_0^{R} r' dr' \int_{-\infty}^{+\infty} dz \frac{\rho_s}{(r_d/r_s)^\alpha(1+r_d/r_s)^{3-\alpha} }
\\
 \\ \noindent \rm{where} \quad  \quad r_d = \sqrt{r'^2 +z^2}
\end{aligned}
\label{eq:Sgnfw}
\end{equation}

The column densities for all these halos were determined from Eq.~\ref{eq:Sgnfw} by numerical integration at  $R=r_s$, where $r_s$ is the scale radius from the gNFW fits. The double integration in the $S_{gNFW}$ function make the standard error propagation intractable, since no closed form analytical solution is available for $S_{gNFW}$. Hence, we have followed an alternative approach to determine the errors. We have calculated the best-fit value and the associated errors in $S_{gNFW}$ from the parameter space of the {\tt emcee} MCMC sampler used for the maximization of the log-likelihood function of the gNFW fits to stacked density profiles. The median values of the aforementioned column-density array give the best value for $S_{gNFW}$ and the 0.16 and 0.84 quantiles, from which one can estimate  1$\sigma$ errors  for $S_{gNFW}$.

\begin{figure*}
    \centering
	\includegraphics[width=0.8\linewidth]{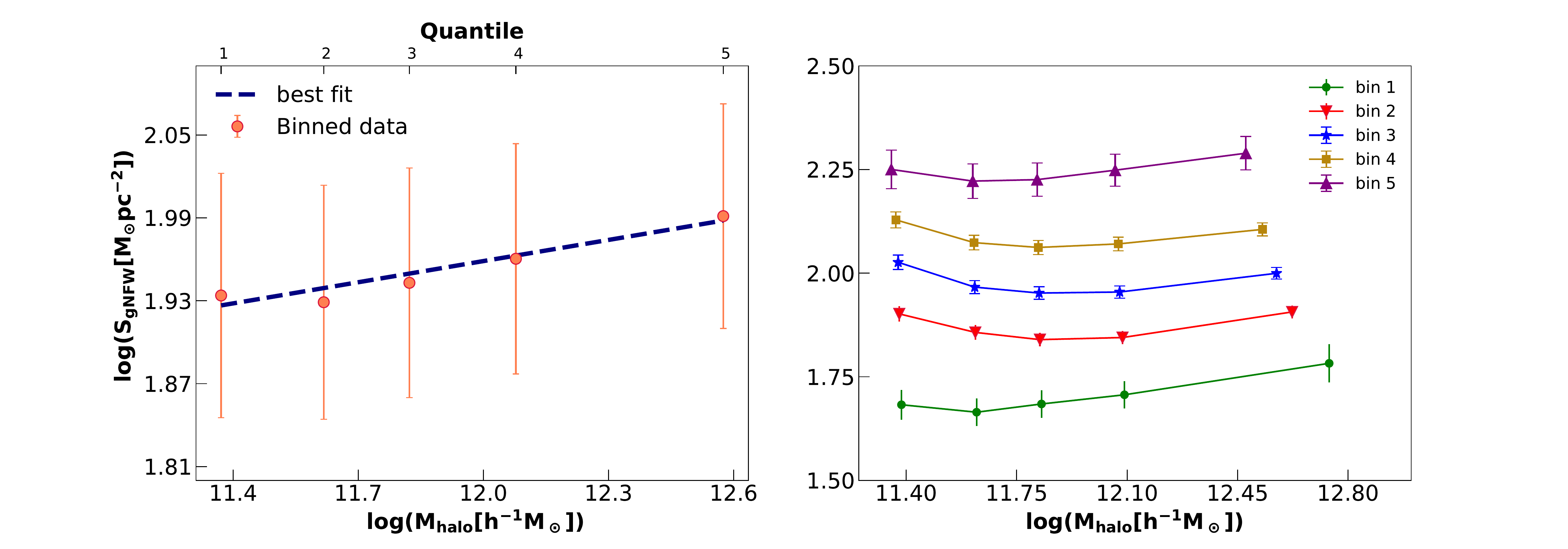}
    \caption{Plot of the dark matter column-density ($S_{gNFW}$) versus halo mass ($M_{halo}$) in log-space. The overall average trend of $S_{gNFW}$ with halo mass from the Set 1 data (left panel) showcases a constant value with the slope equal to $0.053 \pm 0.09$ from a linear-fit. The trend of $S_{gNFW}$ with log $M_{halo}$ and $C_{vir}$  are shown with $C_{vir}$ bins designated with different colours (right panel).}
    \label{fig:figure5}
\end{figure*}

This plot of  $S_{gNFW}$  as a function of halo mass can be found in Figure ~\ref{fig:figure5}.  In the right panel, the  variation of $S_{gNFW}$ with  halo-mass for the Set 2 data is shown with each $C_{vir}$ bin being represented by a different colour.  We can see  that there is a positive  trend between $C_{vir}$ and $S_{gNFW}$ for the same halo mass. Similarly, the plot in  the left panel shows the $S_{gNFW}$ values for the Set 1 data, which have been  calculated in five equal quantiles of mass bins as a function of  halo mass. 
We infer that $S_{gNFW}$ has at best a   weak dependence on $M_{halo}$, but definitely shows  a strong dependence on $C_{vir}$.  

The dark matter column density (after averaging over all concentrations) (Figure ~\ref{fig:figure5})  is given by,
\begin{equation}
    S_{gNFW} \propto M_{halo}^{0.053 \pm 0.09
} 
    \label{eq:Smass}
\end{equation}
Therefore, given the current errors,   $S_{gNFW}$ agrees with both  a  constant value  as well as a power law scaling with  halo mass.
Since all the previous studies on the  halo surface density have used either a cored or NFW profile, a direct comparison of the  absolute value of halo surface density with our simulation results cannot be  easily done.

\section{Conclusions}
\label{sec:conclusions}
A large number of studies have shown that the dark matter halo surface density is constant for a whole suite of galaxies from dwarf galaxies to giant elliptical galaxies over 18 decades in blue magnitude. This invariance of the halo surface density however does not hold for galaxy clusters and galaxy groups. 

In order to ascertain  if these predictions are consistent   with  $\Lambda$CDM, we have investigated the density profiles of mock catalogs obtained using baryonified Gravity-only $\Lambda$CDM simulations, which incorporate  the quasi-adiabatic relaxation of dark matter halos in the presence of baryons~\citep{Paranjapemock}. We tried to fit these halos  with a variety of  cored and non-cored density profiles. Although, the cored isothermal or Burkert profiles can adequately model the rotation curves of galaxies, groups, and clusters~\citep{Salucci19,Kormendy14,Chan},  these cored profiles as well as the cuspy NFW profiles fail to describe density profiles of the mock catalogs, unlike the observed data. Therefore we cannot do a direct detailed comparison to observational results done previously, since they either assumed cored profiles~\citep{Donato,DelPopolo20} or NFW profiles~\citep{Boyarsky}. The simulated halos could only be self-consistently fitted using gNFW profiles over the full range of mass scales. As proxy for the halo surface density, we calculated the dark matter column density for the gNFW profile (cf. Eq.~\ref{eq:eqgnfw}). 

Our conclusions based on these studies are as follows:
\begin{itemize}
    \item The gNFW analytic model best describes the density profiles of the mock simulation catalogs used in this paper. This is very evident from their reduced $\chi^2$ analysis and  model comparison tests such as AIC and BIC, which confirm the fact that the cored  or the NFW profiles do not provide a good fit to these mock catalog objects. 
    \item The gNFW $r_s$ shows a tight positive correlation with halo mass and also a slight decrease in the virial concentration as a  function of halo mass which is similar to that for the standard NFW halos. Furthermore, the gNFW $r_s$ values from the fits to the simulated profiles do not show much deviation from the simulation values of $r_s(NFW)$.  
    \item The dark matter column-density is consistent with \rthis{both a constant halo surface density as a function of halo mass as well as   a power-law dependence with halo mass.}
    \item We also found that $S_{gNFW}$  is positively correlated with $C_{vir}$ for a fixed halo mass.  It would be interesting to compare this with observations.

\end{itemize}
In a follow-up work we also plan to extend these studies to cluster and group-scale halos to  ascertain if the simulations predict the elevated values for the column density at these scales.

\section*{Acknowledgements}
The research of AP is supported by the Associateship Scheme of ICTP, Trieste.  GK acknowledges the Ministry of Education (MoE), Government of India for the Senior Research Fellowship.

\section*{Data Availability}

The data used in this work is described in ~\citet{Paranjapemock} and can be obtained from a reasonable request to the corresponding author of that paper.



\bibliographystyle{mnras}
\bibliography{example} 




\appendix

\section{gNFW fits to Stacked data}
\label{sec:appendix}
In this section we shall discuss about fitting the gNFW profiles to the mass-binned stacks from  Set 1 data. The gNFW model fitting was done by maximising the log-likelihood function given below:

\begin{equation}
\begin{aligned}
\ln \mathcal{L} = -0.5\sum \bigg[\bigg(\frac{\rho_{gNFW}-\rho_{data}}{\sigma} \bigg)^2 + \ln(\sigma^2)\bigg], 
\end{aligned}
\label{eq:maxlike}
\end{equation}
where $\rho_{data}$ and $\sigma$ denote the mock density profiles and their errors, respectively. We apply standard priors to the three free parameters, $r_s$, $\rho_s$ and $\alpha$ of the gNFW model.  A log-uniform prior was employed for both $r_s$ and $\rho_s$ in the range $0<r_s<10^3$ and $10^{-4}<\rho_s<10^{-1}$ respectively.  For $\alpha$, we choose a uniform prior between 0.8 and 1.6. For the MCMC runs, the number of walkers was set to 200 and the total number of iterations to 5000 which attained a mean acceptance fraction of  approximately 0.6 for these fits. 

Figure \ref{fig:appendix} shows the MCMC corner plots showing the posterior distributions with 68\%, 90\% and 99
\% credible intervals of the three free parameters of the gNFW model, viz $\log r_s$,  $\log \rho_s$, and  $\alpha$, which were used for the Set-1 data. The posterior distribution medians and their 1-$\sigma$ uncertainties are represented by vertical dashed lines in the histograms shown diagonally in the corner plots. The corner plots associated with Set 1 data are displayed in ascending order of mass with Figure \ref{fig:a.1} showing the lowest mass bin and  Figure \ref{fig:a.5}, the highest mass bin.

\begin{figure*}
     \centering
     \begin{subfigure}[b]{0.48\textwidth}
         \centering
         \includegraphics[width=\textwidth]{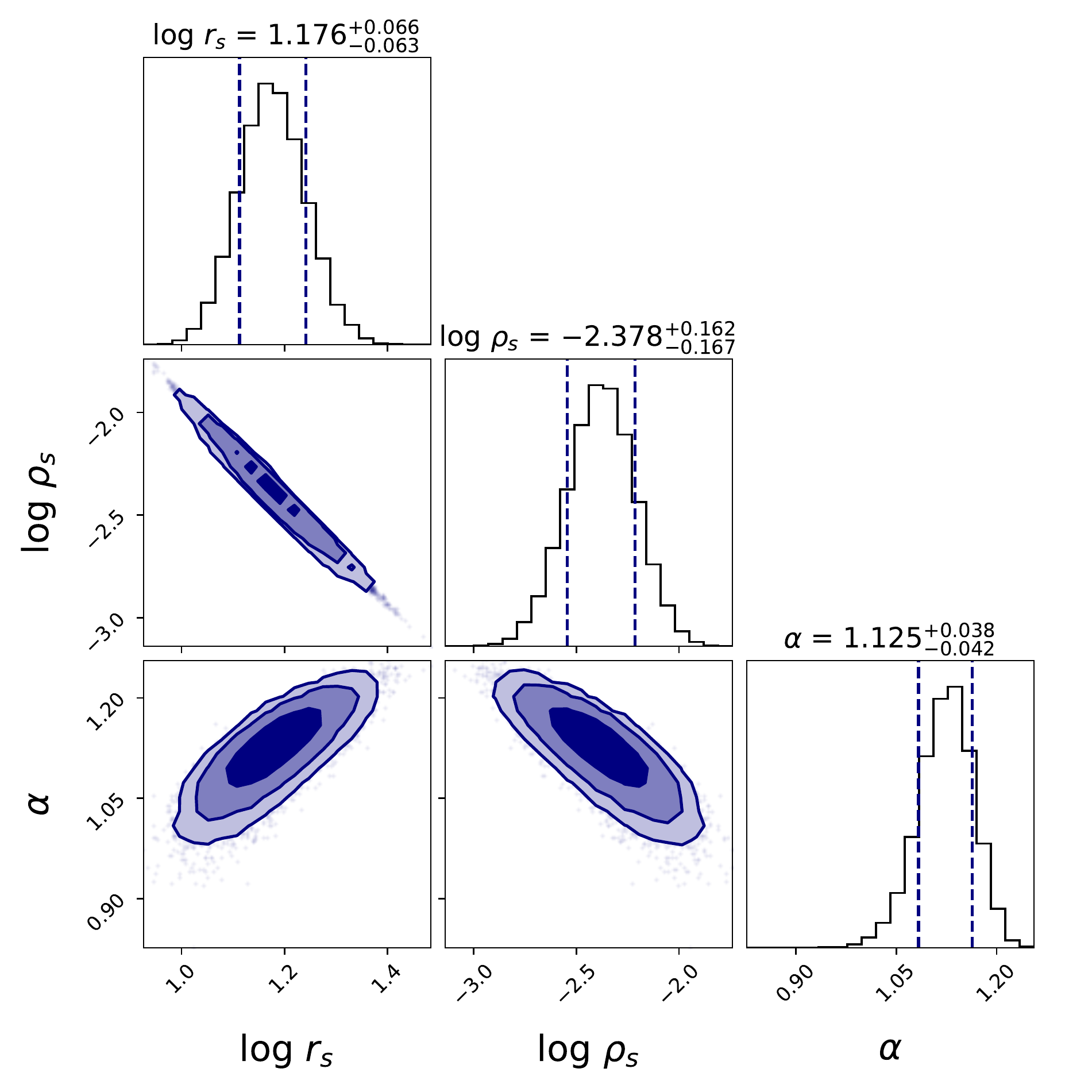}
         \subcaption{Mass bin 1}
         \label{fig:a.1}

     \end{subfigure}
     \hfill
     \hfill
     \begin{subfigure}[b]{0.48\textwidth}
         \centering
         \includegraphics[width=\textwidth]{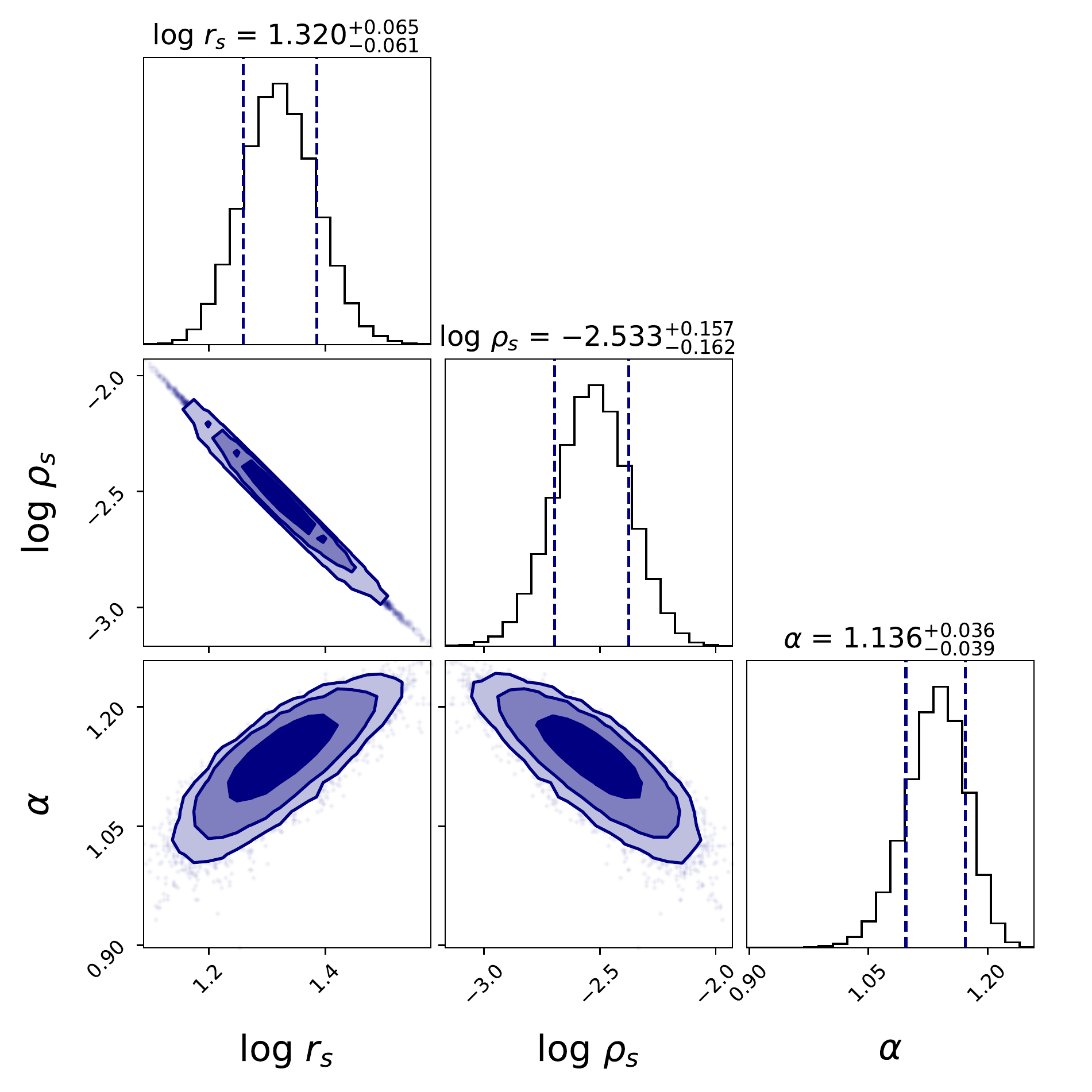}
         \subcaption{Mass bin 2}
         \label{fig:a.2}
     \end{subfigure}
          \hfill
          \hfill

     \begin{subfigure}[b]{0.48\textwidth}
         \centering
         \includegraphics[width=\textwidth]{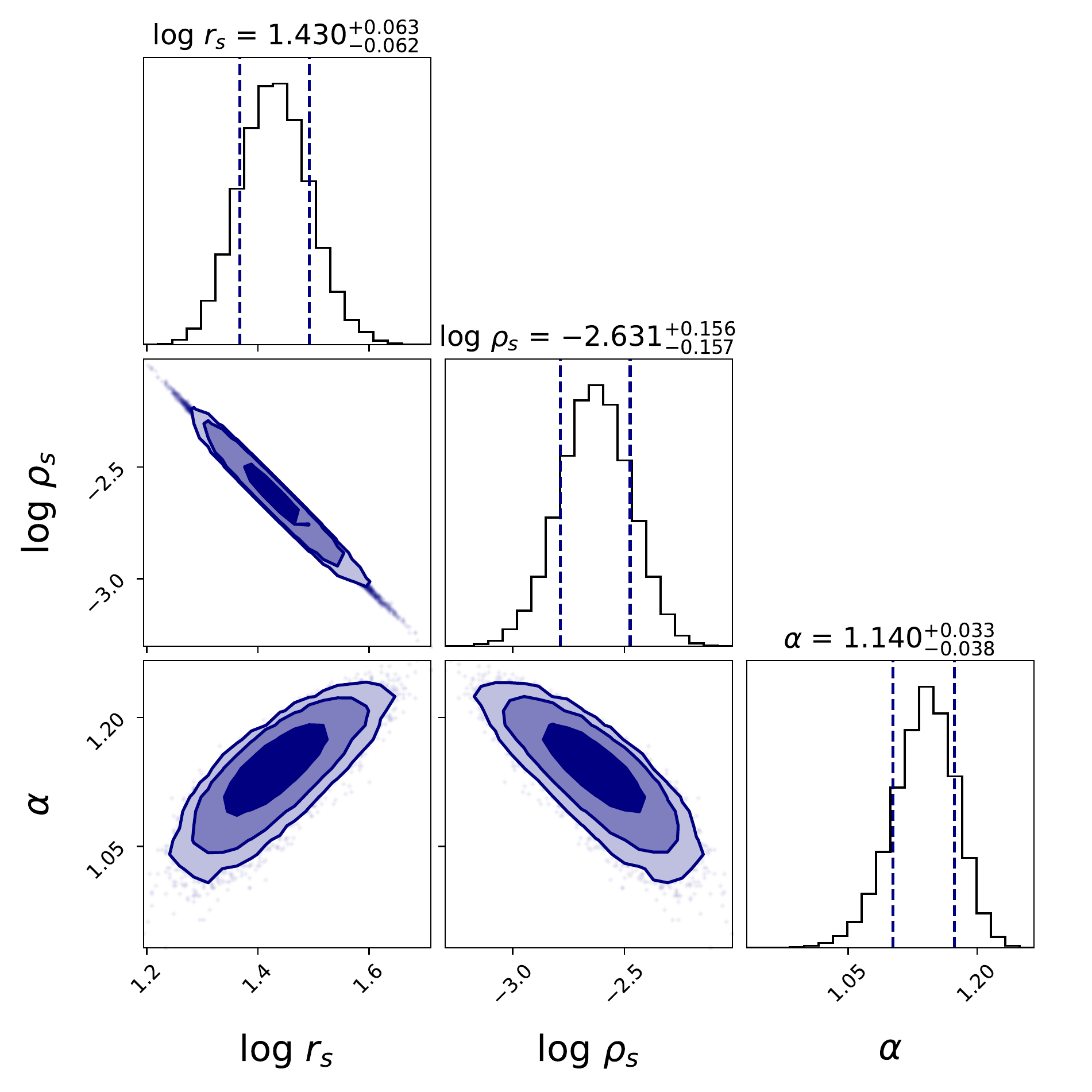}
     \subcaption{Mass bin 3}
         \label{fig:a.3}
     \end{subfigure}
     \hfill
     \hfill
     \begin{subfigure}[b]{0.48\textwidth}
         \centering
         \includegraphics[width=\textwidth]{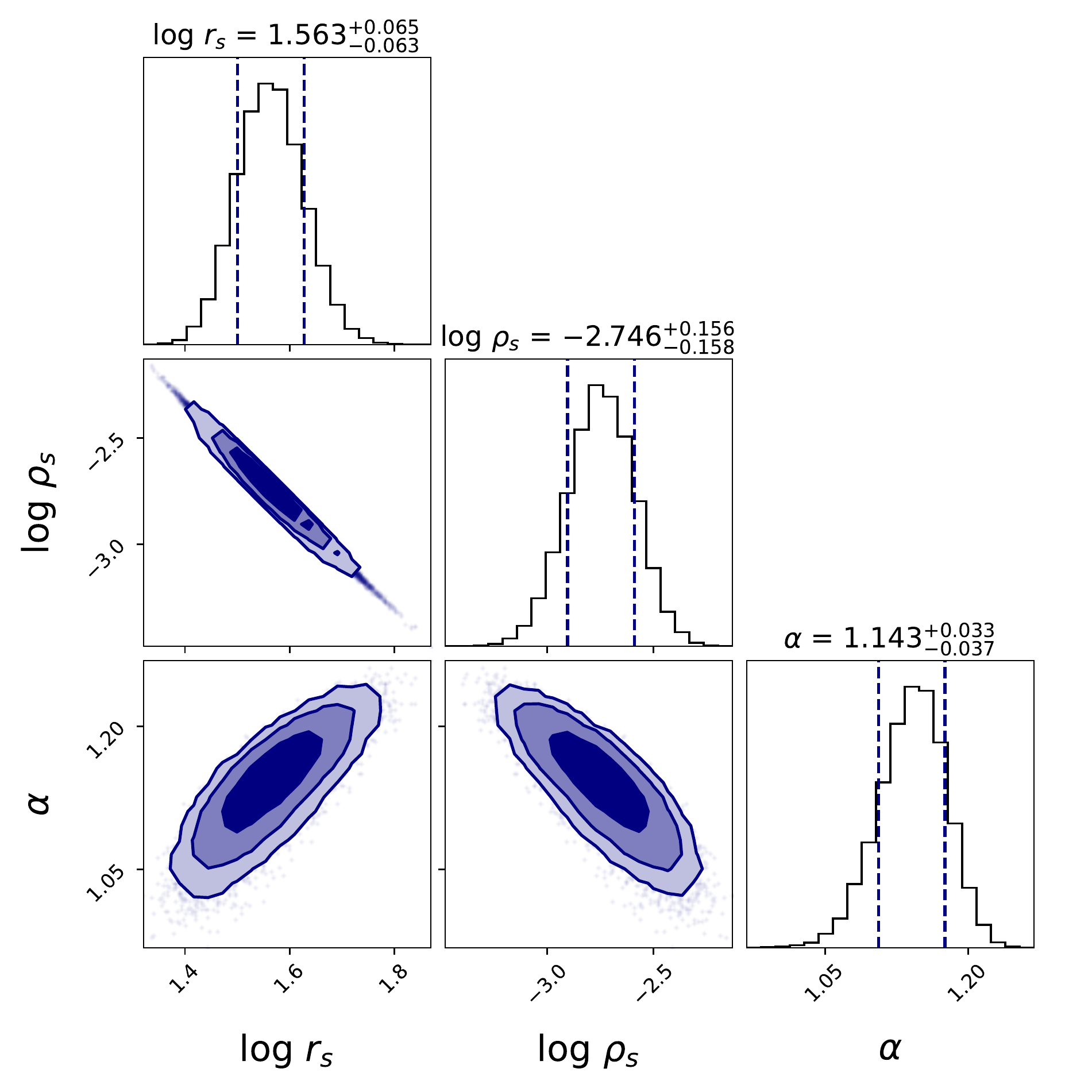}
      \subcaption{Mass bin 4}
         \label{fig:a.4}
     \end{subfigure}

             \caption{Corner plots for the gNFW fits to Set 1 data. The plot shows the 68\%, 90\%, and 99\% credible intervals.}

        \label{fig:appendix}
\end{figure*}

\begin{figure}\ContinuedFloat
    \centering
    \begin{subfigure}[b]{0.48\textwidth}
        \includegraphics[width=\textwidth]{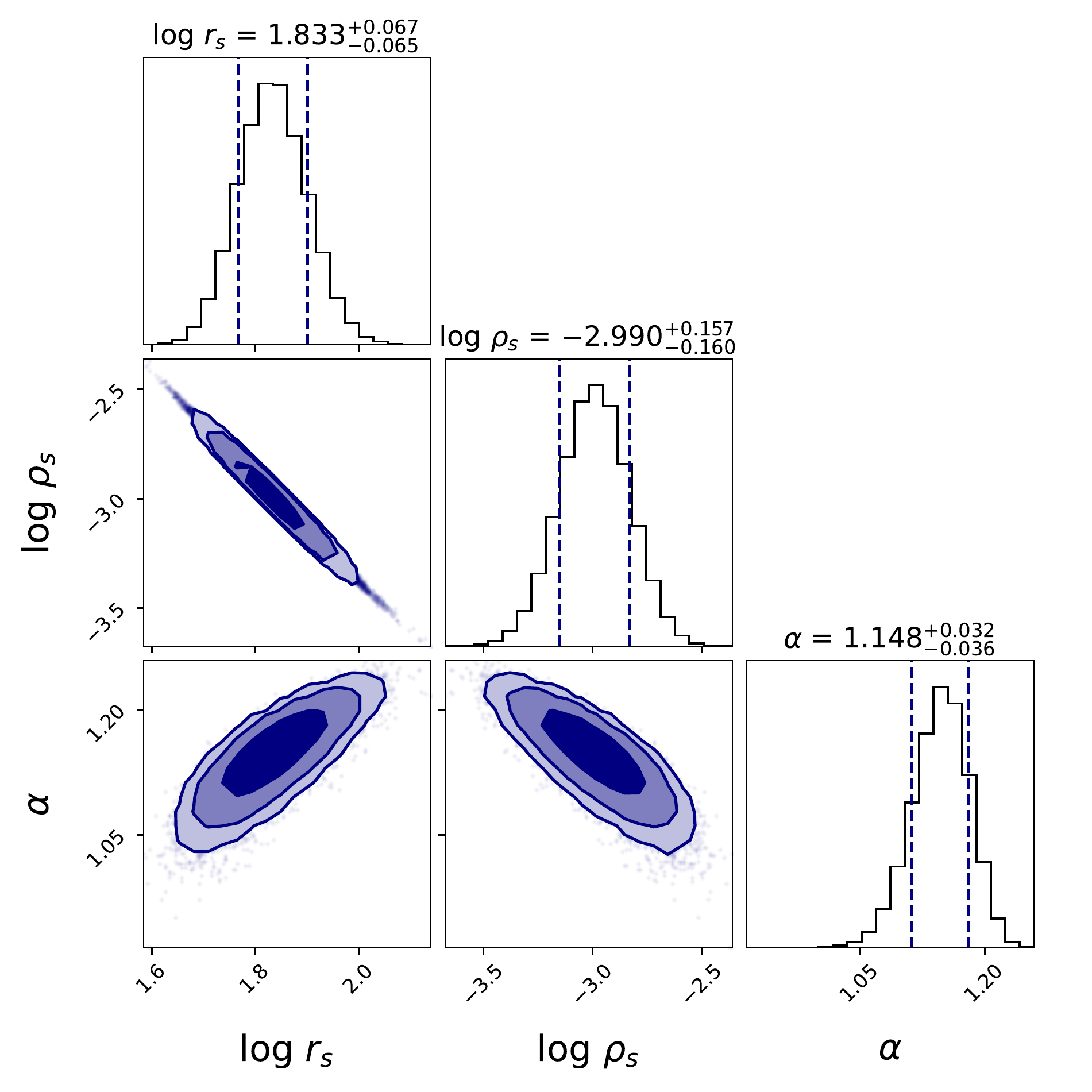}
        \subcaption{Mass bin 5}
        \label{fig:a.5}
    \end{subfigure}

    \caption[]{Corner plots for the gNFW fits to Set 1 data. The plot shows the  68\%, 90\%, and 99\% credible intervals. (cont.)}
\end{figure}


\label{lastpage}
\end{document}